\documentclass{vgtc}                          

\ifpdf
  \pdfoutput=1\relax                   
  \usepackage{graphicx}                
  \DeclareGraphicsExtensions{.pdf,.png,.jpg,.jpeg} 
\else
  \usepackage{graphicx}                
  \DeclareGraphicsExtensions{.eps}     
\fi%

\graphicspath{{figures/}{pictures/}{images/}{./}} 

\usepackage{microtype}                 
\PassOptionsToPackage{warn}{textcomp}  
\usepackage{textcomp}                  
\usepackage{mathptmx}                  
\usepackage{algorithm}                 
\usepackage{algpseudocode}             
\usepackage{times}                     
\usepackage{cite}                      
\usepackage{tabu}                      
\usepackage{booktabs}                  
\usepackage{soul}
\usepackage{algorithm}                 
\usepackage{algpseudocode}  
\usepackage[para,flushmargin]{footmisc}
\usepackage{ragged2e}
\usepackage{etoolbox}
\patchcmd\makefootnoteparagraph{\par}{\RaggedRight\par}{}{}

\newcommand{\pheading}[1]{\vspace{4px}\noindent\textbf{#1}}

\newenvironment{tight_itemize}{\begin{itemize} \itemsep
-1.5pt}{\end{itemize}}

\newcommand{\oscar}{\textsc{Oscar}}

\onlineid{0}

\vgtccategory{Research}

\vgtcinsertpkg

\usepackage{tikz}
\usetikzlibrary{positioning}

\usepackage{subcaption}
\newcommand{\citemissing}[1]{[?]}

\makeatletter
\AtBeginDocument{\let\hl\@firstofone}
\makeatother

\title{\oscar: A Semantic-based Data Binning Approach}

\author{Vidya Setlur\thanks{vsetlur@tableau.com} %
\and Michael Correll\thanks{mcorrell@tableau.com} %
\and Sarah Battersby\thanks{sbattersby@tableau.com}}
\affiliation{\scriptsize Tableau Research}

\teaser{
  \centering
  \begin{subfigure}[t]{0.95\textwidth}
    \centering
     \includegraphics[width=0.49\textwidth]{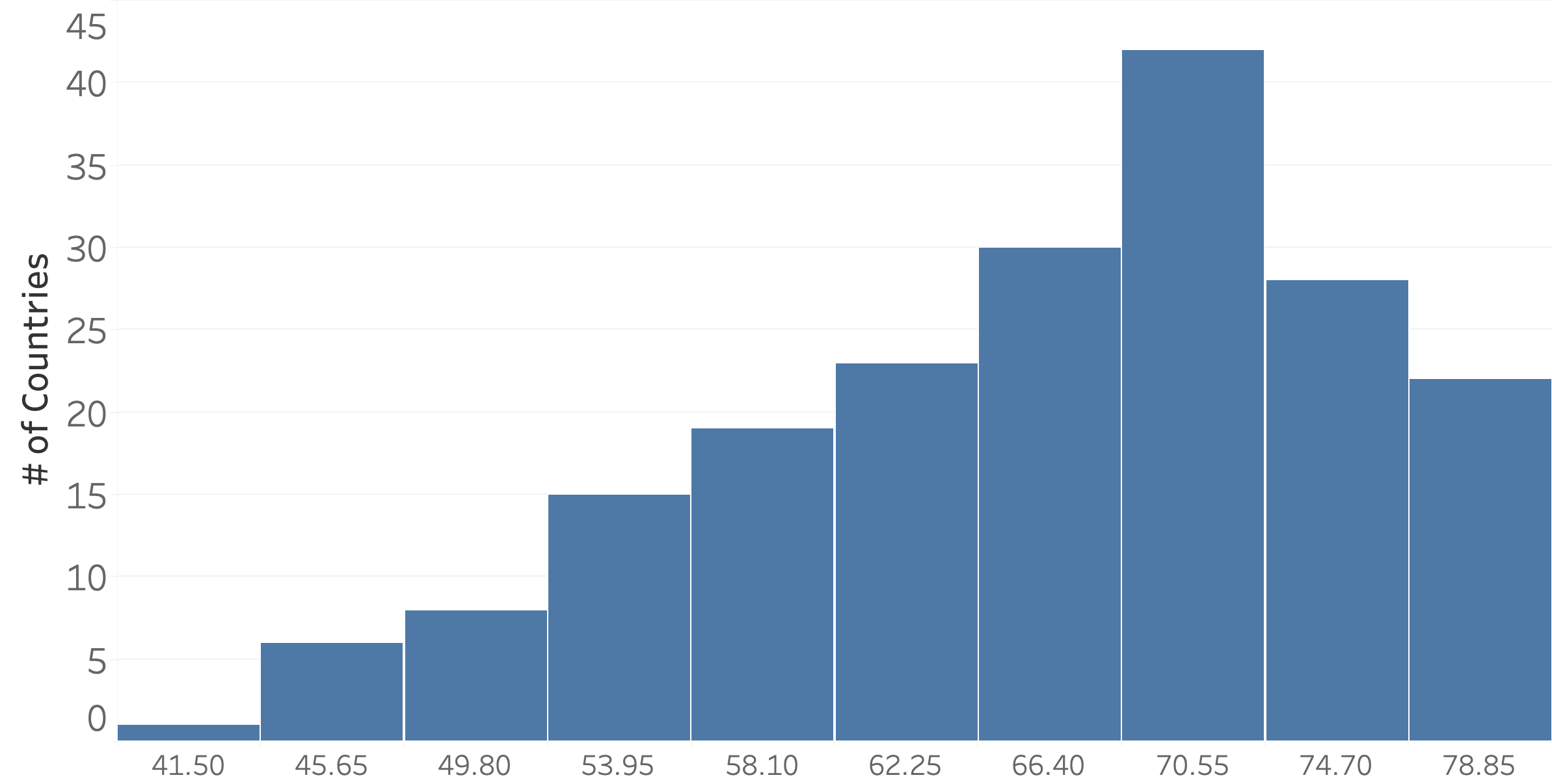}
      \includegraphics[width=0.49\textwidth]{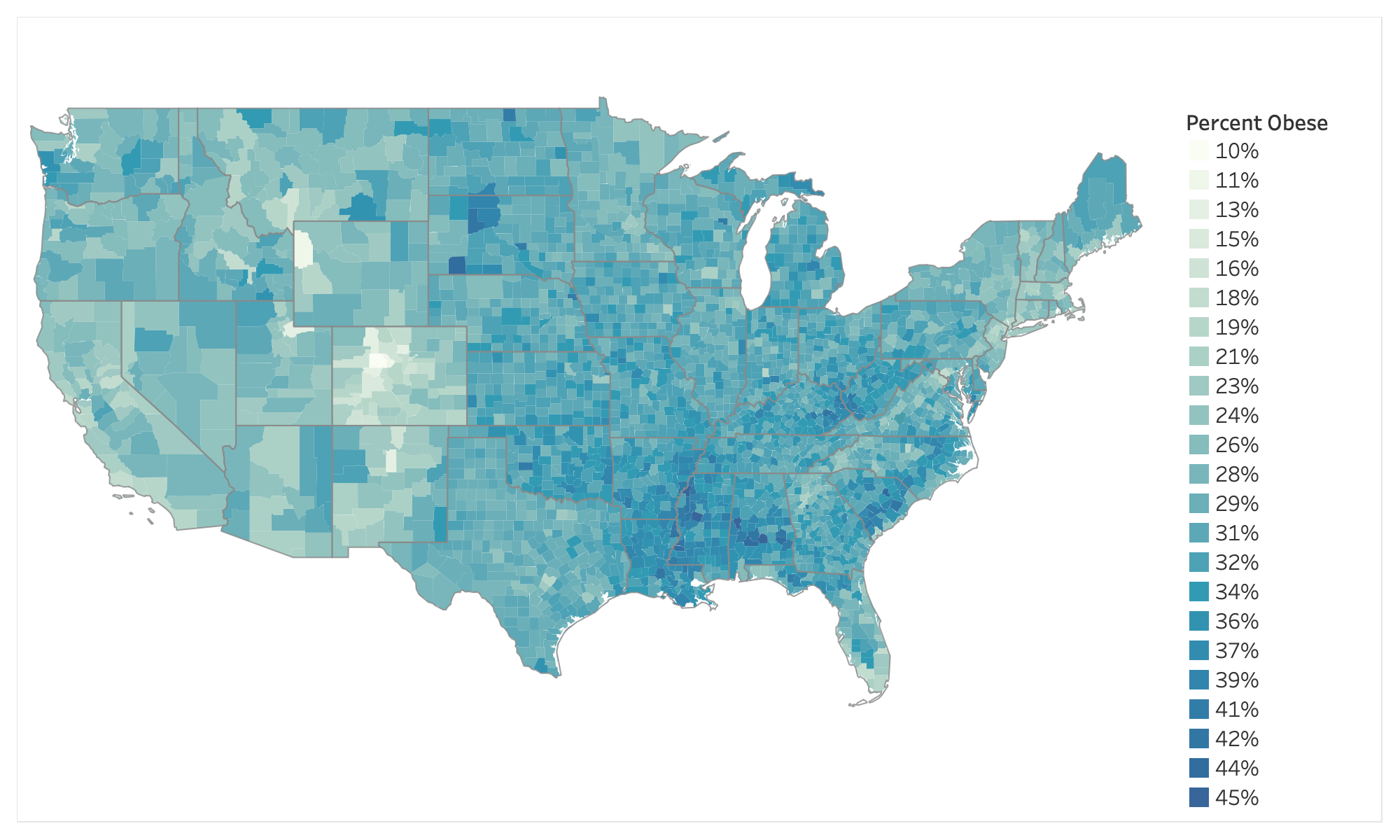}
     \caption{Default binning scheme.}
     \label{fig:teaser-default}
  \end{subfigure}
  
    \begin{subfigure}[t]{0.95\textwidth}
    \centering
     \includegraphics[width=0.49\textwidth]{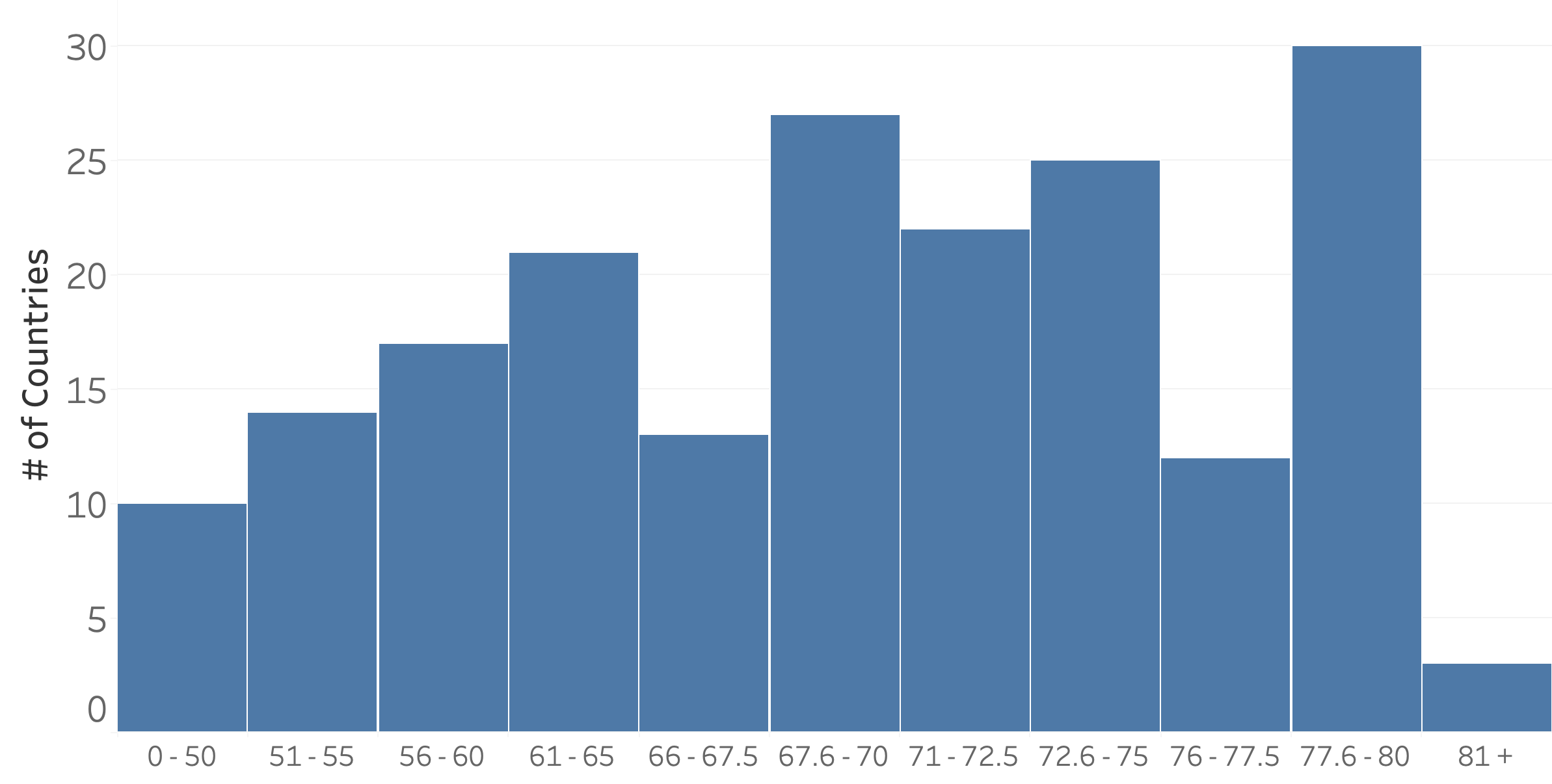}
      \includegraphics[width=0.49\textwidth]{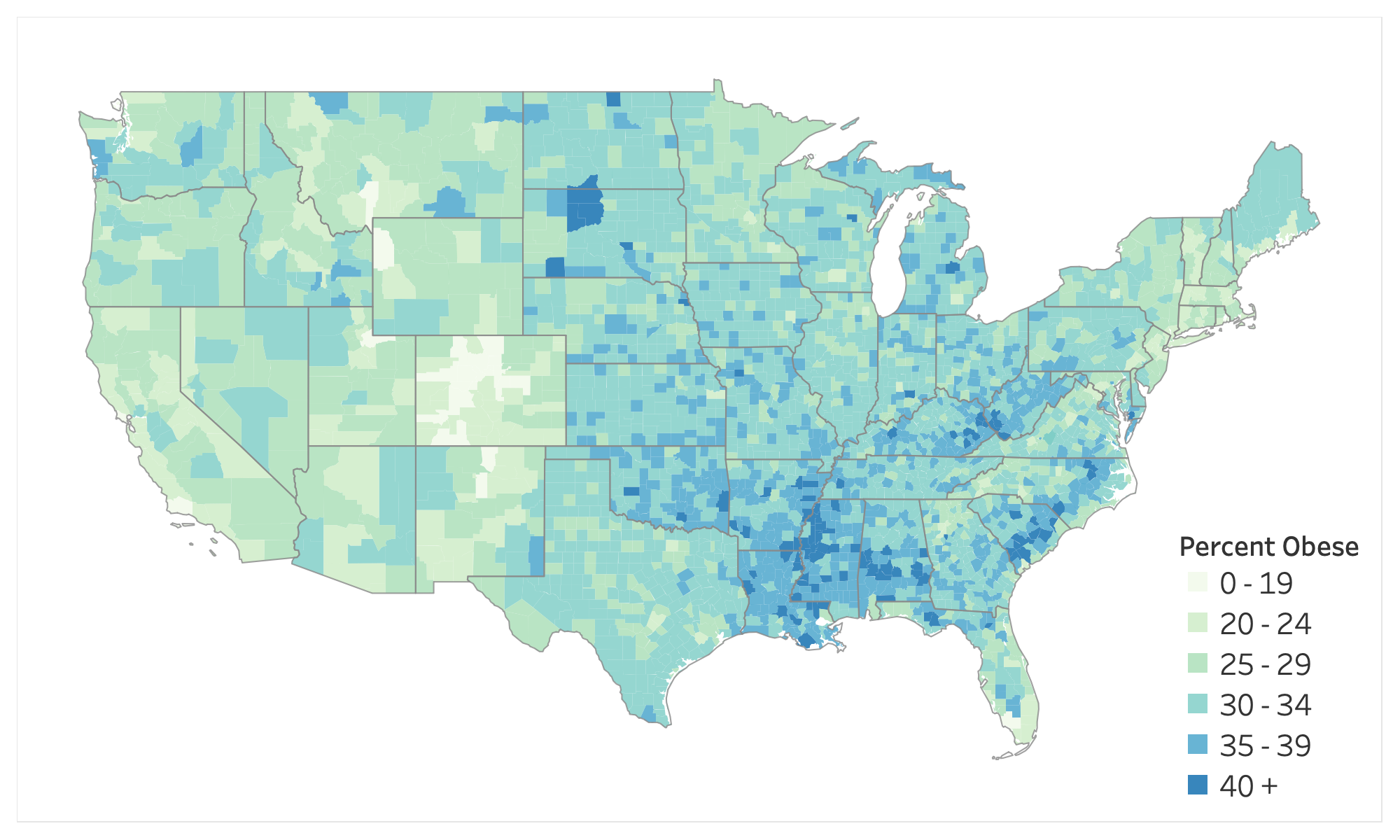}
     \caption{Binning scheme recommended by \oscar}
     \label{fig:teaser-oscar}
  \end{subfigure}
  \caption{
  Visualizations showing comparisons of bins for data on per-country \textit{life expectancy} (left) and per-U.S. county \textit{obesity} rates (right). The top-row bins are computed based on statistical properties, while the bottom-row bins are computed by \oscar. Semantic bins have benefits for legibility, reducing the number of bins (i.e., the visual complexity of the map or histogram), and taking advantage of non-uniformity to either highlight areas of interest or compress long tails of the distribution into single bins.}
  \label{fig:teaser}
}

\abstract{
Binning is applied to categorize data values or to see distributions of data. Existing binning algorithms often rely on \textit{statistical} properties of data. However, there are \textit{semantic} considerations for selecting appropriate binning schemes. Surveys, for instance, gather respondent data for demographic-related questions such as age, salary, number of employees, etc., that are bucketed into defined semantic categories. In this paper, we leverage common semantic categories from survey data and Tableau Public visualizations to identify a set of semantic binning categories. We employ these semantic binning categories in \oscar: a method for automatically selecting bins based on the inferred semantic type of the field. We conducted a crowdsourced study with $120$ participants to better understand user preferences for bins generated by \oscar~vs. binning provided in Tableau. We find that maps and histograms using binned values generated by \oscar~are preferred by users as compared to binning schemes based purely on the statistical properties of the data.
} 


\CCScatlist{
  \CCScatTwelve{Human-centered computing}{Visualization}{}
}

\begin{document}


\firstsection{Introduction}
\maketitle
Binning \hl{of quantitative attribute data} is a prerequisite step for many aspects of data visualization--- the bins can be used to reduce continuous data down to more manageable categories, preserve data privacy through aggregation, generate histograms, or create breaks for ordinal color scales. Different binning schemes prioritize different goals: for instance, the selection of the number of histogram bins may be chosen to minimize error compared to an unknown but estimated population distribution~\cite{freedman1981}, or the bins used for a color palette set in order to maximize both intra-bin coherence and inter-bin difference~\cite{jenks1977optimal}. Yet, these schemes based on statistical properties can ignore what might be the most crucial property of a binning scheme meant for use in visualization: the \textit{legibility and semantic coherence of the bins}. The ultimate consumers of these binning schemes are human readers of charts and maps, and so a human-centered binning scheme ought to leverage not just the \textit{statistics} of the quantitative field in question but also \textit{data semantics} as well as the \textit{legibility and interpretability} of the resulting bins.

In this paper, we present \oscar~\footnote{The name \oscar~is inspired by the beloved Sesame Street character who embraces his life living out of a trash \emph{bin}~\cite{oscar}.}, a human-centered binning technique that leverages data semantics and legibility constraints to suggest bins for quantitative data for use in histograms, maps, and other charts. \oscar~leverages information from Tableau Public~\cite{public} to suggest common bin sizes for a particular field based on the field's name (or those of fields with semantically similar names), then allows the user to refine the resulting bins. The resulting bins provide a number of useful features for legibility (\autoref{fig:teaser}), including a focus on particular values of interest for specific use cases, a respect for the grain of the data, and the use of non-uniform bins to condense long tails or outliers into single bins. \hl{\protect\oscar{ } addresses attribute "binning" or classification; it does not perform any spatial binning.}

We conduct a crowdsourced evaluation of the bins generated from \oscar~ and find a general preference for ``semantic'' bins over bins created via software defaults such as Tableau's~\cite{calculations}. Beyond our binning scheme, findings also suggest generalizable principles for ``human-centered'' binning, such as user preferences for ``nice'' bin boundaries (e.g., whole numbers or numbers rounded to multiples of 5 or 10) and bin boundaries with appropriate granularities that do not collapse information about critical semantic values (e.g., having finer-grained bins for higher age values for \textit{life expectancy}). In addition to specific recommendations for creating human-legible bins, preliminary evidence indicates that \oscar{ }provides useful bin semantics for creating charts that better match the needs of users.

\section{Related Work}

\subsection{Histogram Binning}
Selecting the correct number of bins in a histogram is often portrayed as a tradeoff--- too few bins and the distinct shape of a distribution is lost; too many, and the resulting noisy histogram makes shape information difficult to recover. A common assumption is also that the quantitative data in a histogram are \textit{samples} from an unknown \textit{population} distribution: the choice of the number of bins is characterized as an estimation problem, and common histogram binning schemes (such as Sturges' rules~\cite{scott2009sturges}, the Freedman-Diaconis rule~\cite{freedman1981}, and Scott's rule~\cite{scott1979optimal}) are based on minimizing error under certain assumptions about the population distribution (e.g., for Sturges' rule, that it is a unimodal Gaussian) and under certain definitions of error (e.g., for Scott's rule, Mean Integrated Squared Error).

However, histograms are generated for diverse audiences and for diverse purposes, and these rules may fall short for different human-scale tasks. For instance, Correll et al.~\cite{correll2019looks} find that common rules may generate too few bins for people to reliably identify data quality issues like missing data in distributions (rather than merely their shape). Conversely, Sahann et al.~\cite{sahann2021perception} find that relatively few bins are sufficient for viewers to reliably distinguish between different population distributions and suggest there are diminishing returns for creating more bins. Lastly, Gopal Lolla et al.~\cite{gopal2011selecting} find that important shape information can be lost within traditionally assigned bins and suggest an error function that incorporates shape information when selecting bin boundaries for histograms. We point to these works to suggest the potential benefits of a \textit{human-centered} binning schema that is mindful of how people read (or, just as importantly, fail to read~\cite{boels2019conceptual,lem2013misinterpretation,kaplan2014student}) histograms and how they are employed for a variety of analytical goals beyond estimating the shape of a distribution.

\subsection{Cartographic binning}

With maps, binning provides an opportunity to explore patterns across spatial distributions. The color encoding on the map enforces visual grouping of regions based on bin category. While continuous or un-classed maps are valuable for maintaining absolute numeric data relationships~\cite{tobler1973choropleth}, it is more common to use discrete bins to emphasize the similarity between locations.  These bins should be meaningful for the dataset \textit{and} the question(s) being explored; they may be driven by data distribution (e.g., standard deviation), be more arbitrary and unrelated to data distribution (e.g., equal interval), or tied to specific visual / data values of relevance to the cartographer or map reader (e.g., diverging bins arranged around the income requirement for a Federal assistance program)~\cite{slocum2014thematic}.  Some of the most commonly provided cartographic binning methods~\cite{brewer2002evaluation} are equal interval, Jenk's optimal or natural breaks~\cite{jenks1977optimal}, mean/standard deviation, quantiles, and ``pretty breaks'' (rounding to nice-looking numbers), though there are generally also options for manual bin range selection providing more opportunity to tailor the view based on user goals or understanding of data distribution.  Less commonly implemented for commercial usage, but still academically interesting to consider are \hl{ automated methods such as those relying on genetic algorithms}~\cite{armstrong2003using}, or proximity-based binning schemes~\cite{monmonier1973maximum} to encourage more spatially compact, homogeneous regionalization on the map. The use of color also places a constraint on the number of bins: MacEachren~\cite{maceachren1982role} has noted that while more detailed maps may be interpretable, the information retained from map reading tends to be reduced to roughly three ordinal categories (high, medium, low).

\subsection{Incorporating semantics into visualization design}

Other techniques have augmented visualization designs through the use of automatic lookups meant to resonate with the semantic backgrounds of viewers. Kim et al.~\cite{kim2016analogies} create personalized analogies based on location data to help users better interpret distances (e.g., explaining concepts in terms of the distance from San Francisco to San Jose might be easier to conceptualize for a Bay Area resident than for the distance from London to Reading). Most similar to our work is that of Setlur \& Stone~\cite{setlur2016linguistic}, who employ a linguistic approach for automatically 
generating semantically resonant color palettes for categorical data. Their algorithm uses Google Image data to find associations between words and colors (e.g., for a bar chart of vegetable produce sales, assigning green and orange colors to the data values `broccoli' and `carrot', respectively). Inspired by this work, we demonstrate a technique for applying external domain knowledge and semantics of common binning patterns and categories to help inform reasonable defaults for histograms and breaks for classes in data. For data attributes that do not have semantic associations, we apply best practice binning techniques to provide default bins.

\section{\oscar~Binning Process}
The algorithm employs a two-step process for creating more useful bins for numerical data attributes: (1) Semantic bins for fields that have semantic lookups and (2) default bins that create human-legible bin breaks based on the underlying statistical properties in the data.
\subsection{Generate Semantic Bin Lookup}
\label{sssec:semanticlookup}
To generate a lookup of semantic categories and their corresponding bins, we employ a data-driven approach of mining both a public corpus of survey questionnaires~\cite{MCSQ} and published Tableau Public~\cite{public} visualizations containing binned fields. \hl{Leveraging a corpus of prevalent semantic categories provides recognizable and familiar bin breaks that are often used in data analysis.} The core idea of our
technique is employing Latent Dirichlet Allocation (LDA)~\cite{lda}, a popular topic modeling technique to extract topics from a given corpus as proxies for semantic bin concepts. 

\pheading{Build the LDA model for bin concepts}. We build an LDA model from the binned field names from Tableau dashboards along with common demographic information commonly found in surveys that have binned numerical responses such as age, salary, population, etc. The LDA model is trained using MALLET~\cite{McCallumMALLET} for $1000$ iterations with the top $100$ binned field names. We then apply the LDA model on the survey corpus to get probabilities for bin concepts in each survey to generate a lookup of strings and their associated bin sizes.

\pheading{Create a list of related concepts}. For each bin concept, we have a label name and a set of related concepts such as synonyms. Enriching the bin concept with related concepts increases the probability of a match with the LDA model. These seed lists are created using a thesaurus service~\cite{thesaurus} and Wordnet synsets~\cite{wordnet}. For example, the bin concept `salary' has a seed list:[ `pay', `payroll', `base salary', `wage', `remuneration', `stipend', `earnings', `income'].

\pheading{Align bin concept and bin breaks}. 
The final step to creating the semantic bin lookup is an alignment between the bin concept $c$, along with its related concept list (together we denote as $R(c)$), and question topic $t$ in the surveys with every topic being mapped to at most one concept. We align each $t$ to $c$ with the maximum score, $S(c,t)$ that measures the summed probabilities of $c$ and $R(c)$ in $t$: $S(c,t) = \sum_{w\in R(c)}p(w|t)$. We remove all alignments where $S(c,t) < a_{threshold}$. In practice, we found that $a_{threshold} = .06$ provides reasonable results for precision and relevance.

\subsection{\oscar~binning algorithm}

\subsubsection{Compute Semantic Bins}

\begin{figure}
  \centering
  \includegraphics[width=\linewidth]{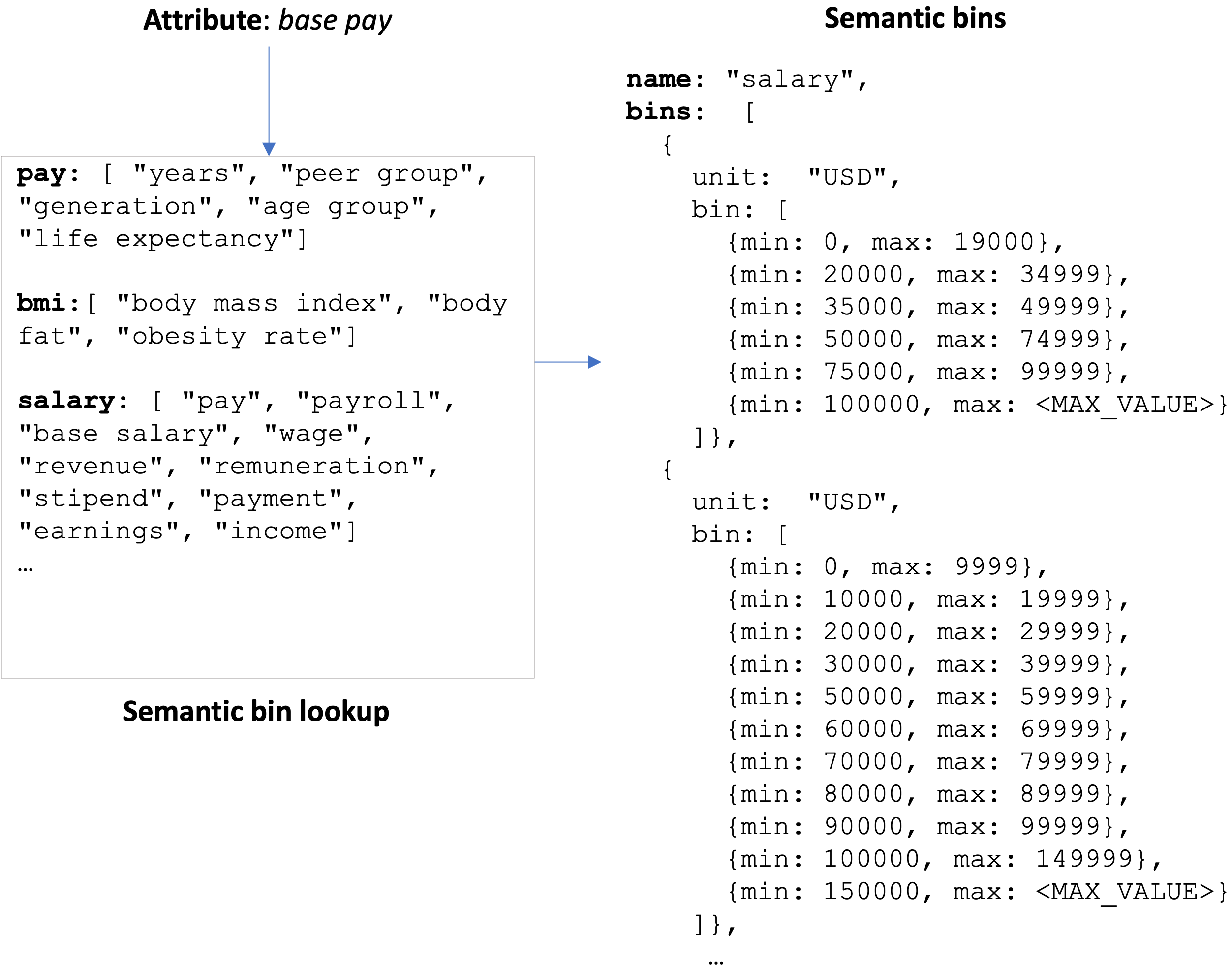}
  \caption{Semantic bin lookup for the attribute \textit{base pay}~in \oscar.}
  \label{fig:lookup}
\end{figure}

Given an attribute, \oscar~first checks if there is a match with the attribute and the semantic bin lookup, as described in Section~\ref{sssec:semanticlookup}. We employ fuzzy matching and lemmatization to match attribute strings to bin concepts in the table~\cite{nlpbook}. If there is a match and there is a semantic bin option, these bins are applied to the attribute to generate the corresponding visualization (e.g., a histogram or a map shown in Figure~\ref{fig:teaser-oscar}). If multiple options of semantic bins are available, the semantic bin option that is closest to the data bounds of the attribute is selected. An overview of this process is illustrated in Figure~\ref{fig:lookup}.

\subsubsection{Compute Default Bins}
\label{sec:refinement}
\begin{figure}[htb]
  \centering
    \begin{subfigure}[t]{0.47\columnwidth}
    \centering
     \includegraphics[width=\textwidth]{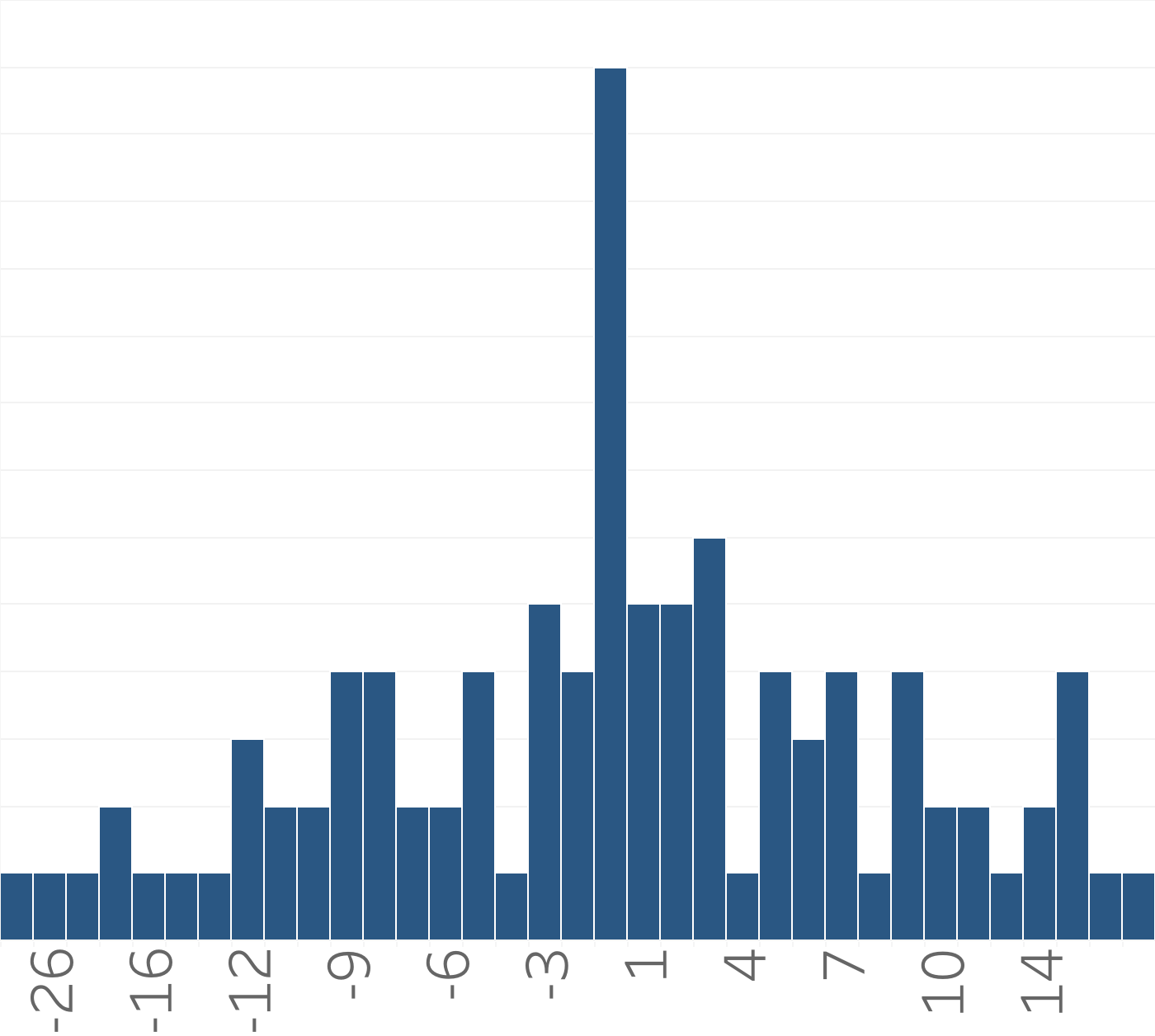}
     \caption{Too many bins}
     \label{fig:goldilocks-toomany}
  \end{subfigure}
  ~
    \begin{subfigure}[t]{0.47\columnwidth}
    \centering
     \includegraphics[width=\textwidth]{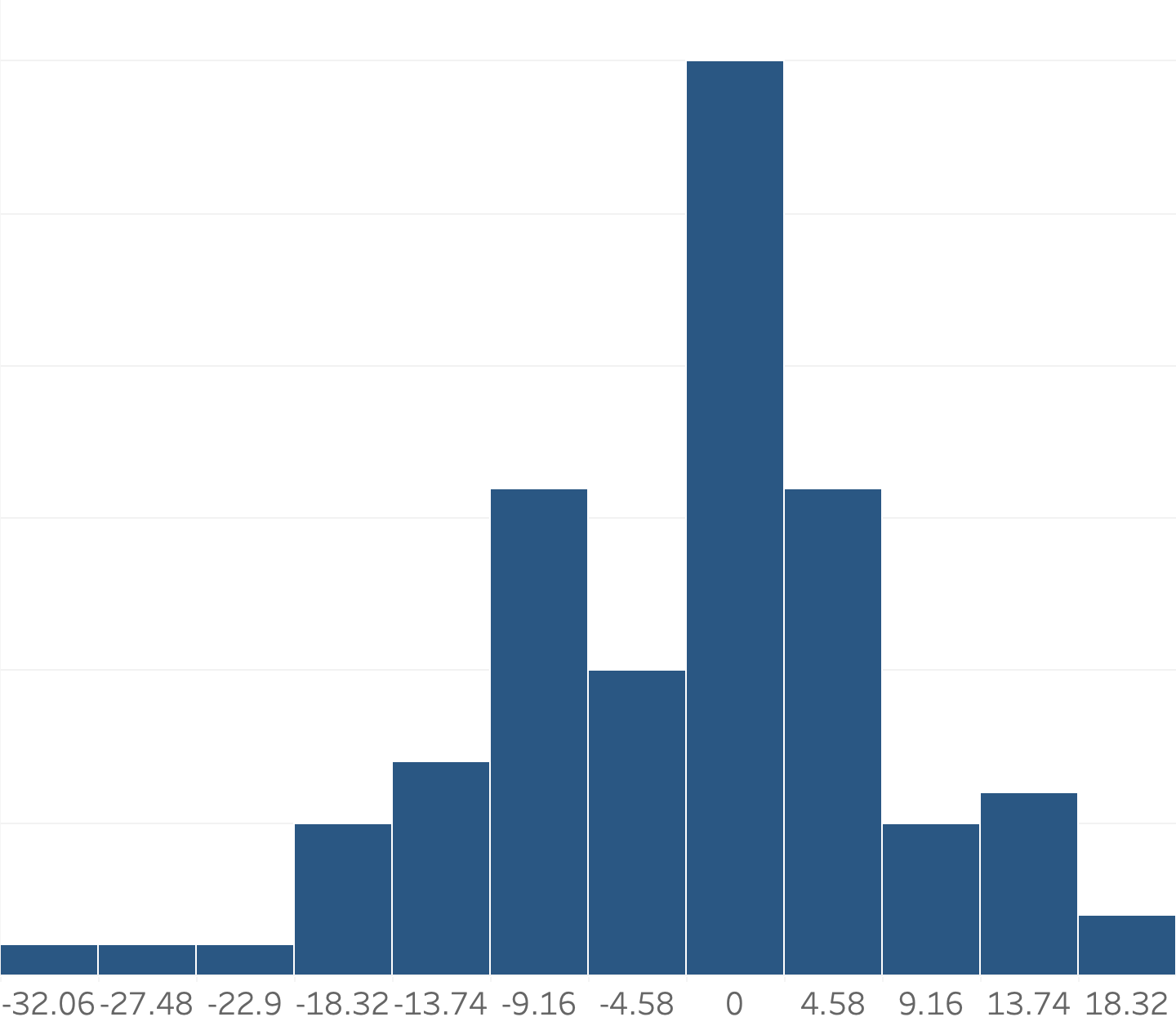}
     \caption{Inappropriate bin precision}
     \label{fig:goldilocks-grain}
  \end{subfigure}
  
    \begin{subfigure}[t]{0.47\columnwidth}
    \centering
     \includegraphics[width=\textwidth]{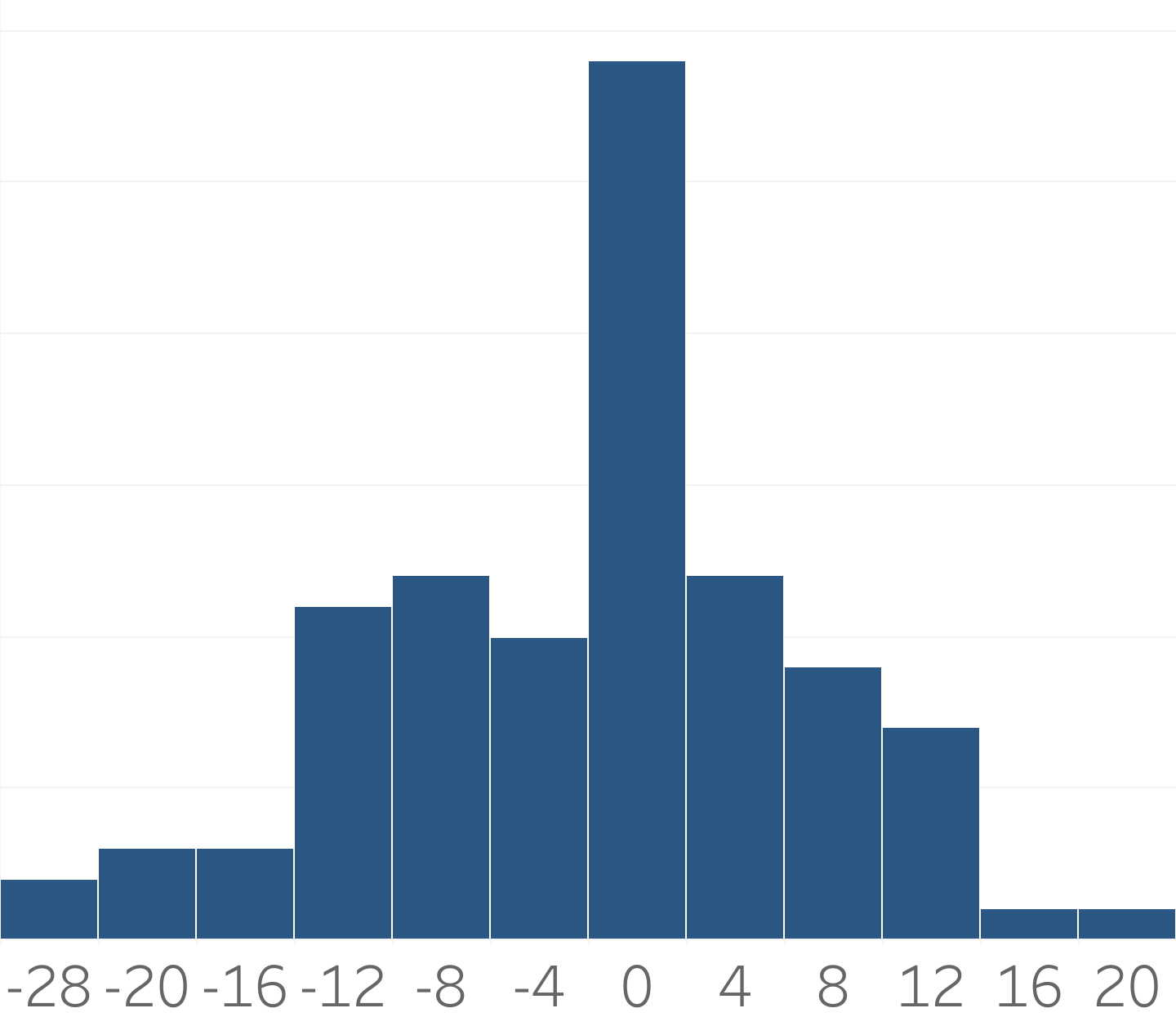}
     \caption{Bins not beginning from nice powers of 5 or 10.}
     \label{fig:goldilocks-notnice}
  \end{subfigure}
  ~
  \begin{subfigure}[t]{0.47\columnwidth}
    \centering
     \includegraphics[width=\textwidth]{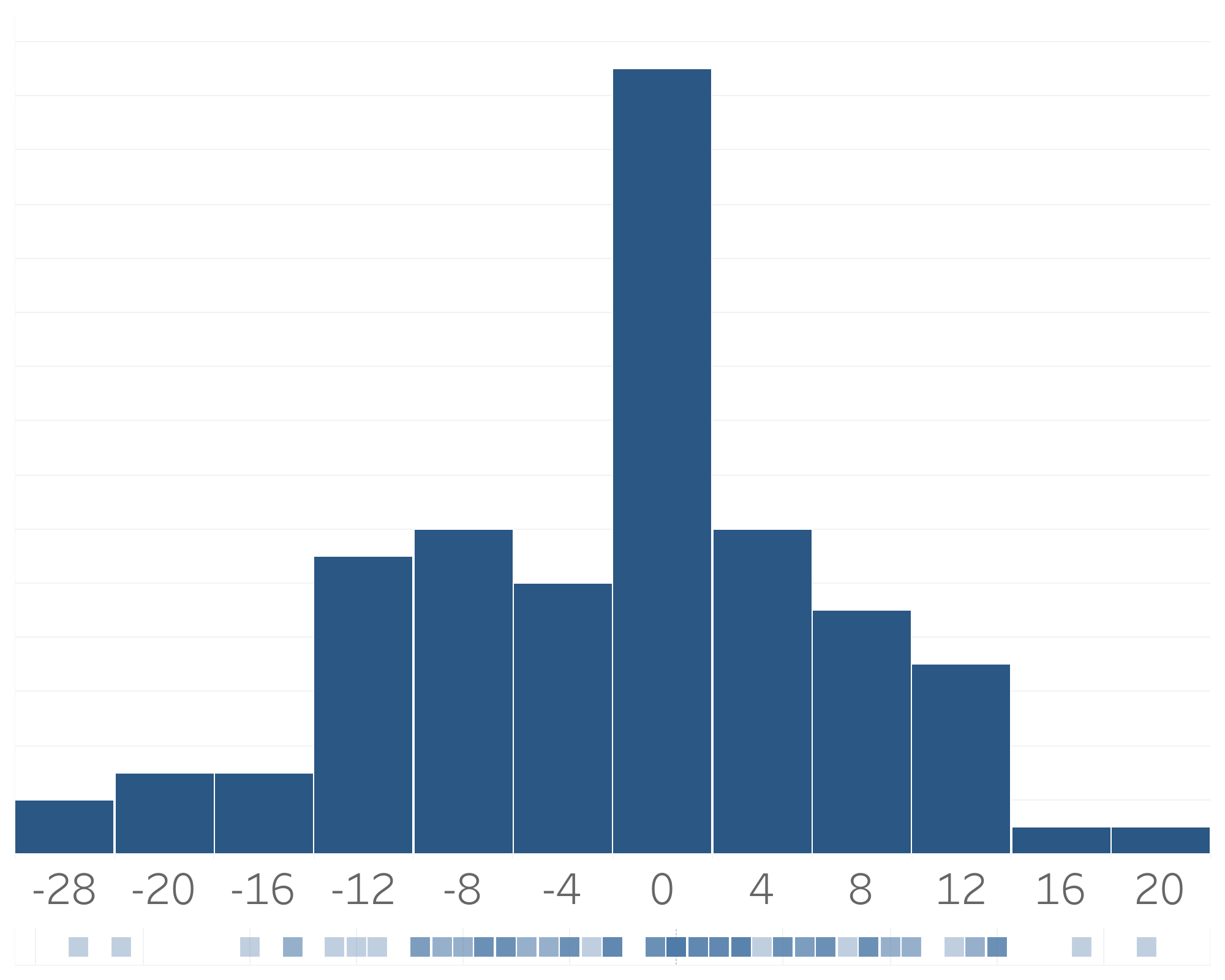}
     \caption{A final binning scheme, with a strip plot of raw values for comparison.}
     \label{fig:goldilocks-good}
  \end{subfigure}
  \caption{
\hl{Default binning} (\protect\autoref{fig:goldilocks-good}) \hl{heuristics applied on a dataset of 100 integers drawn from a normal distribution. The bins are 1) limited in number (especially for use in color ramps, where the ability to distinguish colors is limited, compared to} \protect\autoref{fig:goldilocks-toomany}), 2) \hl{consistent with the data grain (e.g., if the data are integers, then the bins should not be decimal numbers as in} \protect\autoref{fig:goldilocks-grain})\hl{, and 3) rounded to ``nice'' values (say, powers of 5 or 10, rather than widths of 4 in \protect\autoref{fig:goldilocks-notnice}).}}
 \label{fig:goldilocks}
\end{figure}

In the absence of semantic bins, \oscar~attempts to use smart defaults to select human-legible bins (Figure \ref{fig:goldilocks}). After choosing a binning based on the statistical properties of the distribution (e.g., Sturges rule~\cite{scott2009sturges}), \oscar~performs the following optimizations:

\pheading{Constrain the number of bins to a maximum number.} For use in a color ramp, the designer might wish to limit the number of bins in order to maximize the discriminability of colors or reduce the complexity of the legend (see Figure \ref{fig:teaser-default}, right for an example of a perhaps overly complex color legend). For use in a histogram, the designer might wish to make sure that there are not so many bins that features in the distribution or the labels of bins are illegible (see Figure \ref{fig:goldilocks-toomany}). While these maximums are to some extent dependent on contingent properties such as display resolution or color ramp choice, we note that the default maximum number of bins in VegaLite~\cite{satyanarayan2017vegalite} is 20, and the maximum number of bins for stepped color ramps in ColorBrewer~\cite{harrower2003brewer} is 12 (and even then, only with a subset of ramps).

\pheading{Round the bin extents to match the precision of the data.} Many existing rules for selecting histogram bins can produce arbitrary precision floating point bin boundaries. These boundaries can be misleading if they promise or suggest precision beyond the precision of the data. For instance, integer data should not have decimal bins (see Figure \ref{fig:goldilocks-grain}), and data expressed in terms of round millions of dollars should not have bin widths of tens of thousands of dollars.

\pheading{Round the bin extents to convenient values of 5 or 10}. We draw inspiration here from VegaLite's~\cite{satyanarayan2017vegalite} ``nice'' operator, which rounds bin or scale extents to an appropriate power of 10 given the precision of the value (e.g., $9\rightarrow 10$, $0.9\rightarrow 1.0$, etc.).

\pheading{Shift bins such that $0$ does not occur within a bin}. Again drawing inspiration from VegaLite~\cite{satyanarayan2017vegalite}'s ``anchor'' operator, we believe there is a semantic distinction between positive and negative numbers for a wide variety of quantitative fields. By shifting bin boundaries, we can ensure that 0 falls between bins.

\vspace{-1em}
\section{Evaluation}
In this study, we explored the high-level research question: What are participants' preferences when viewing charts with and without semantic binning? We hypothesized the following: 
\begin{tight_itemize}
\item \textbf{H1}: Participants will prefer semantic bins over default bins to better reflect the categories for the corresponding data attribute.
\item \textbf{H2}: Participants will not prefer coarse semantic bins as it may be harder to discern distribution patterns.
\item \textbf{H3}: Participants will not prefer non-integer bin starts in the default bins as these bins are harder to interpret.
\end{tight_itemize}

\subsection{Participants}
We conducted a power analysis to determine the number of participants required to find an overall difference in preference rankings. With a medium effect size of $\sim$0.49, our analysis suggests that a target sample of 112 would yield 95\% power to detect an overall difference between preference rankings at an alpha level of 0.05. 

We recruited 125 participants from Amazon Mechanical Turk. To qualify for the study, participants were required to be located in the USA and have a 95\% acceptance rate on previous tasks. We compensated participants at a rate of \$2.00 for a six-minute study. After excluding participants who failed attention checks (e.g., failing to select a specific answer in a multiple-choice question) or entered nonsensical responses, we ended up with 120 participants, with 64 that identified as women ($M_{age}$ = 34.05, $SD_{age}$ = 10.52), 52 as men ($M_{age}$ = 36.65, $SD_{age}$ = 11.88), and four chose to not disclose.  

The participants completed a subjective graph literacy report~\cite{garcia2013communicating} and reported an average value of 3.85 out 6 ($SD = 0.82$, 1 = low self-reported literacy, 6 = high self-reported literacy), suggesting that most participants were comfortable with visualizations but did not identify as visualization experts. Only 7 people reported that they create visualizations often for work or as a hobby, and 21 people reported rarely interacting with visualizations in their daily lives. 

\subsection{Stimulus and Design}
We created two sets of ranking stimuli - one set using \oscar, each differing in the number of bins ($5-14$ bin categories) and the other based on Tableau binning functionality~\cite{tableauhistogram}. For generating these bins, we used five datasets: CDC obesity health data~\cite{obesity}, World Indicators~\cite{gapminder}, US Census commute data~\cite{census}, country Gini coefficient~\cite{gini}, and Titanic passenger data~\cite{titanic}. Six sets of histograms and four sets of maps were generated, resulting in a total of ten preference tasks. We denote the \oscar~binned histograms and maps as $Histo_{oscar1}$, $Histo_{oscar2}$, $Map_{oscar1}$, and $Map_{oscar2}$, with $oscar1$ having less number of bins than $oscar2$; the default binned histogram and map as $Histo_{default}$ and $Map_{default}$ respectively.

\subsection{Procedure}
Participants were provided a link and a brief introduction to our survey. Participants viewed all the ten preference tasks in random order. For each task, participants were shown the default and semantic binning variants of chart images and were asked, ``Which chart is more useful for showing the distribution of the data.'' They were provided a free-text response field to add feedback about their ranking choices. At the end of the survey, participants reported demographic information and completed the self-report visual literacy test. 

\subsection{Analysis}
We conducted a Friedman Rank Sum test using the PMCMRplus R package~\cite{pohlert2018package} to compare preference rankings for the visualizations, with post-hoc pair-wise comparisons via Conover's test with Bonferroni's correction to determine the ranking differences.  

\subsection{Results}
 There is a significant difference in user preferences semantically binned charts overall, \hl{supporting} \textbf{H1}. For the histogram charts, we observed $Histo_{oscar1}$ over $Histo_{default}$ (Friedman $\chi^2$ = 34.62, $p < 0.001$), $Histo_{oscar2}$ over $Histo_{default}$ (Friedman $\chi^2$ = 20.59, $p < 0.001$), and $Histo_{oscar2}$ over $Histo_{oscar1}$ (Friedman $\chi^2$ = 19.36, $p < 0.001$). For maps we observed $Map_{oscar1}$ over $Map_{default}$ (Friedman $\chi^2$ = 54.25, $p < 0.001$), $Map_{oscar2}$ over $Map_{default}$ (Friedman $\chi^2$ = 68.32, $p < 0.001$), and $Map_{oscar2}$ over $Map_{oscar1}$ (Friedman $\chi^2$ = 69.33, $p < 0.001$).
 
Observations confirm that users prefer finer-grained semantic bins over coarser-grained ones (\textbf{H2}). Also, low preference rankings for default binned charts containing non-integer bin starts \hl{support} \textbf{H3}.
 
 Feedback from the participants also reflected these observations. $P27$ stated, ``The obesity numbers seemed to be a bit arbitrary (referring to the default bins) and wanted the images that showed something I'm familiar with.'' Other comments on bin granularity included, ``I don't like the histogram clumping up everything. Prefer to see it all spread out [$P76$]'' and ``I don't see that much color variation with the smaller categories. Liked seeing more differences [$P07$].'' There was also feedback against non-integer bin breaks - ``What do those decimals even mean? I can't wrap my head around them [$P71$]'' and ``It's easy on the eyes to see the whole numbers; they are less scary compared to all those floating numbers [$P100$].''

\section{Discussion and Future Work}

By relying on prior bins constructed by users of Tableau Public that are likely to have semantic importance, as well as by affording defaults that are oriented around the human legibility and interpretability of bins, \oscar~was able to produce binning schemes that were often preferred over binning schemes that are only sensitive to the distribution. We view these initial successes as evidence that the automatic integration of semantic information from usage data can improve visualization design and that the careful setting of defaults can influence the success or failure of a visualization design. However, we see several areas for further improvement and study, both in terms of the design and evaluation \oscar~as well as for human-centered visualization design, especially for mass audiences. 

\pheading{Provide user affordances for bin refinement and repair}. Our semantic binning schemes could result in poor outcomes for a variety of reasons, such as poorly chosen defaults or concept drift as data semantics change over time (e.g., an early map of COVID case data would need to be re-binned as the scale of the pandemic changed). To counteract these potential failures or shifts, We envision \oscar~to take more of a mixed-initiative role in dealing with underspecification, allowing users to ``repair''~\cite{setlur2019under} instances of under-specification, or directly compare different binning schemes (as in Figure \ref{fig:teaser}), \hl{or to tailor their binning schema to particular contexts (for instance, putting different constraints on bins to be used for generating color ramps, as opposed to bins used to generate histograms)}. 

\pheading{Evaluate \hl{quality of} semantic bins \hl{for various} analytical tasks}. \hl{The semantic bin lookup method employs a corpus-driven approach that identifies bin breaks prevalent for various commonly occurring concepts such as salary, population, and age. While the LDA model provides an effective approach for identifying probable semantic categories with corresponding bins, further investigation needs to explore the quality of these bins based on user intent and the analytical task at hand.} Our evaluation is preliminary and focused on \hl{user} preference. It remains future work to investigate if there could be a preference-performance gap for certain tasks (such as identifying missing data or modes)~\cite{correll2019looks}, or determine if there are quantifiable benefits to semantic bins beyond their legibility.

\section{Conclusion}
This paper presents a technique, \oscar~that automatically selects bins based on the inferred semantic type of the data attribute. Using a combination of data-driven semantic lookup information obtained from public survey corpora and Tableau dashboards containing binned fields, \oscar~provides semantic bins and smart defaults to generate human-legible bins. We conducted a crowdsourced user preference study with $120$ participants to better understand user preferences for bins generated by \oscar~vs. default binning provided in Tableau. We find that maps and histograms using binned values generated by \oscar~are preferred by users as compared to binning schemes based purely on the statistical properties of the data. These preliminary results indicate that \oscar~provides useful bin semantics that could be incorporated into visual analysis workflows to create more semantically meaningful charts.

\newpage
\bibliographystyle{abbrv-doi}

\bibliography{bibtex}
\end{document}